Pulsed Laser Deposition of $La_{0.67}Ca_{0.33}MnO_3$ thin films on $LiNbO_3$

Andreas Heinrich[1], Andreas L. Hörner[2], Achim Wixforth[2] and Bernd Stritzker[1]

[1]Institute of Physics, Experimentalphysics IV, University Augsburg, 86135 Augsburg, Germany

[2]Institute of Physics, Experimentalphysics I, University Augsburg, 86135 Augsburg, Germany

Abstract

Surface Acoustic Waves on piezoelectric substrates can be used to investigate the dynamic conductivity of thin films in a non-contact and very sensitive way, especially at low conductivities. Here, we report on such surface acoustic wave studies to characterize thin manganite film like $La_{0.67}Ca_{0.33}MnO_3$, exhibiting a Jan Teller effect with a strong electron phonon interaction and a metal insulator transition at high temperatures.

We report on the deposition of $La_{0.67}Ca_{0.33}MnO_3$ on piezoelectric substrates ($LiNbO_3$ in different crystal cuts) employing a pulsed laser deposition technique. The structural qualities of the thin films are examined by X-Ray Diffraction, Scanning Electron Microscope and Energy Dispersive X-ray spectroscopy. For the electrical characterization, we employ the surface acoustic wave technique, accompanied by conventional DC-resistance measurements for comparison.

## 1. Introduction

Manganites like $La_{0.67}Ca_{0.33}MnO_3$ (LCMO) exhibit a Jan-Teller effect with a strong electron phonon interaction [1] and a metal insulator transition (MIT) at high temperatures (above 200K). To experimentally investigate the MIT, or more general the dynamic conductivity of thin films, surface acoustic waves (SAW) turned out to be an ideal tool [2]. This is because a SAW on a piezoelectric substrate is accompanied by piezoelectric fields propagating at the speed of sound, which strongly interact with mobile carriers in the thin film. On piezoelectric substrates, like $LiNbO_3$ (LNB), SAW are readily excited by means of so-called interdigital transducers (IDT) by injection of a HF signal into such comblike interdigitated metal electrodes on top of the surface.

Within this work we deposited LCMO thin films with our pulsed laser deposition (PLD) system on (100) $SrTiO_3$, (100) MgO YZ-cut and 128°cut $LiNbO_3$ substrates. In the following we will very briefly describe our Laser Ablation System and give details on the film deposition. X-ray diffraction data and SEM images will be discussed as well as the resulting DC resistance curves. For comparison, contact less SAW transmission data will be presented and discussed, accordingly.

## 2. Experimental details

### Laser Ablation System

A typical laser ablation system consists of a vacuum recipient with a heated sample holder, and a target of the material to be deposited onto the substrate. This target is irradiated through a window by an intense laser beam, generating a flash of the target material towards the substrate. Details of the laser ablation system in general are described elsewhere [2].

In our setup, we used 30 ns UV-pulses ($\lambda$=248 nm), generated by an excimer laser (KrF / LPX 300 / Lambda Physics). A lens in front of the recipient chamber projects an aperture onto the rotating target. The cross section of the beam is thereby reduced about 10 times and hence the energy density on the target is concentrated to 3.5-10 J/cm$^2$. This results in an evaporation of the target material and a plasma plume perpendicular to the target surface. During the ablation process oxygen pressure and laser energy are adjusted to let the plasma plume just reach the sample.

Film deposition

Table I gives an overview of the structural parameter of the used substrates (SrTiO$_3$, MgO, LiNbO$_3$) and of LCMO. As one can see, there is always a strong lattice mismatch between the substrate and the LCMO. In the case of YZ-cut LiNbO$_3$ and 128°-cut LiNbO$_3$ , the (300) or the (104) plane is parallel to the surface, respectively.

| Material | a [nm] | b [nm] | c [nm] | $\varepsilon = \dfrac{a_x}{a_{LCMO}}$ | $\varepsilon = \dfrac{b_x}{b_{LCMO}}$ | $\varepsilon = \dfrac{b_x}{c_{LCMO}}$ |
|---|---|---|---|---|---|---|
| (100) SrTiO$_3$ | 3.91 | 3.91 | 3.91 | 0.72 | 0.51 | 0.71 |
| LiNbO$_3$ | 5.15 | 5.15 | 13.86 | 0.94 | 0.67 | 0.94 |
| YZ-LiNbO$_3$ | 5.15 | 13.86 | - | 0.94 | 1.8 | 2.53 |
| 128°-LiNbO$_3$ | 5.15 | 6.21 | - | 0.94 | 0.81 | 1.14 |
| La$_{0.67}$Ca$_{0.33}$MnO$_3$ | 5.45 | 7.70 | 5.47 | - | - | - |

Table I: overview of the structural parameter of the used substrates (SrTiO$_3$, MgO, LiNbO$_3$) and of LCMO.

For the actual LCMO deposition, an oxygen background pressure of p$_O$=1 mbar, a heater temperature of T$_H$=800°C and a laser repetition rate of f$_{rep}$=10Hz was used (in general 10.000 Pulses, resulting in a film thickness of about d=400nm). The energy density in this case can be estimated to be about D$_E$=4J/cm$^2$.

The film morphology was then determined by Scanning Electron Microscope (SEM at 15 kV). The crystallographic structure was determined by X-Ray Diffraction (XRD) (Cu$_{K\alpha}$) and the film composition by Energy Dispersive X-ray spectroscopy (EDX).

After structural characterisation, the LCMO films are laterally patterned into a Hall-bar geometry employing a lithographical structured photo resist etch-mask, followed by a wet etching step in HCl. In a second lithographical step, contact pads for the Hall-bar and the interdigital-transducers (IDT) are defined by optical lithography, metal-deposition and a following lift-off step. We used IDTs with a wave length of $\lambda_{SAW}$=26.5 µm and N=42 pairs of fingers. The aperture of the IDT and hence the width of the acoustic path was chosen to be $W_{SAW}$= 604 µm = 23 $\lambda_{SAW}$. Perpendicular to the SAW beam of the acoustic delay line, a thin LCMO Hall bar (2mm x 15µm) with additional Ohmic contacts for DC 4-probe-measurements is defined (Fig. 1).

The DC Hall bar measurements where performed with a constant current source and a voltmeter. For the SAW- transmission measurements an standard SAW setup consisting of a RF-generator and a Spectrum Analyzer was used. All low temperature measurements were done in a time resolved zero span mode, and in Helium atmosphere. For this purpose, the sample was placed in a magnet cryostat containing a variable temperature insert (VTI) with a temperature range between T= 4.2 and T= 300 K. The cryostat can provide magnetic fields normal to the sample surface with a maximum magnetic field of $B_{max}$= 9 T.

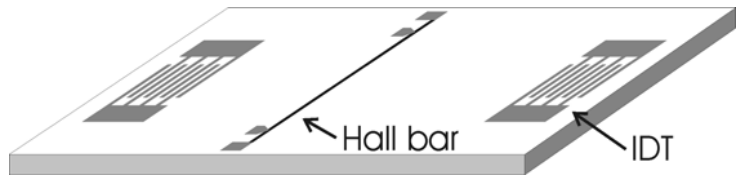

Fig. 1: Chip-design with two IDTs and LCMO hall-bar used for DC and SAW measurements. The thin Hall bar like LCMO stripe in the center of the SAW delay line is the sample under investigation.

## 3. Results and discussion

<u>Structural and crystallographic characterization</u>

After successful deposition, the topography and structural morphology of the films was investigated. In Fig. 2, we show typical SEM images of a LCMO film grown on $SrTiO_3$ (a), $128°$-$LiNbO_3$ (b), YZ-$LiNbO_3$ (c). In the case of $SrTiO_3$ a smoother grain structure, apart from droplets from the target, can be recognized. In comparison to that LCMO on $LiNbO_3$ exhibits a very rough surface.

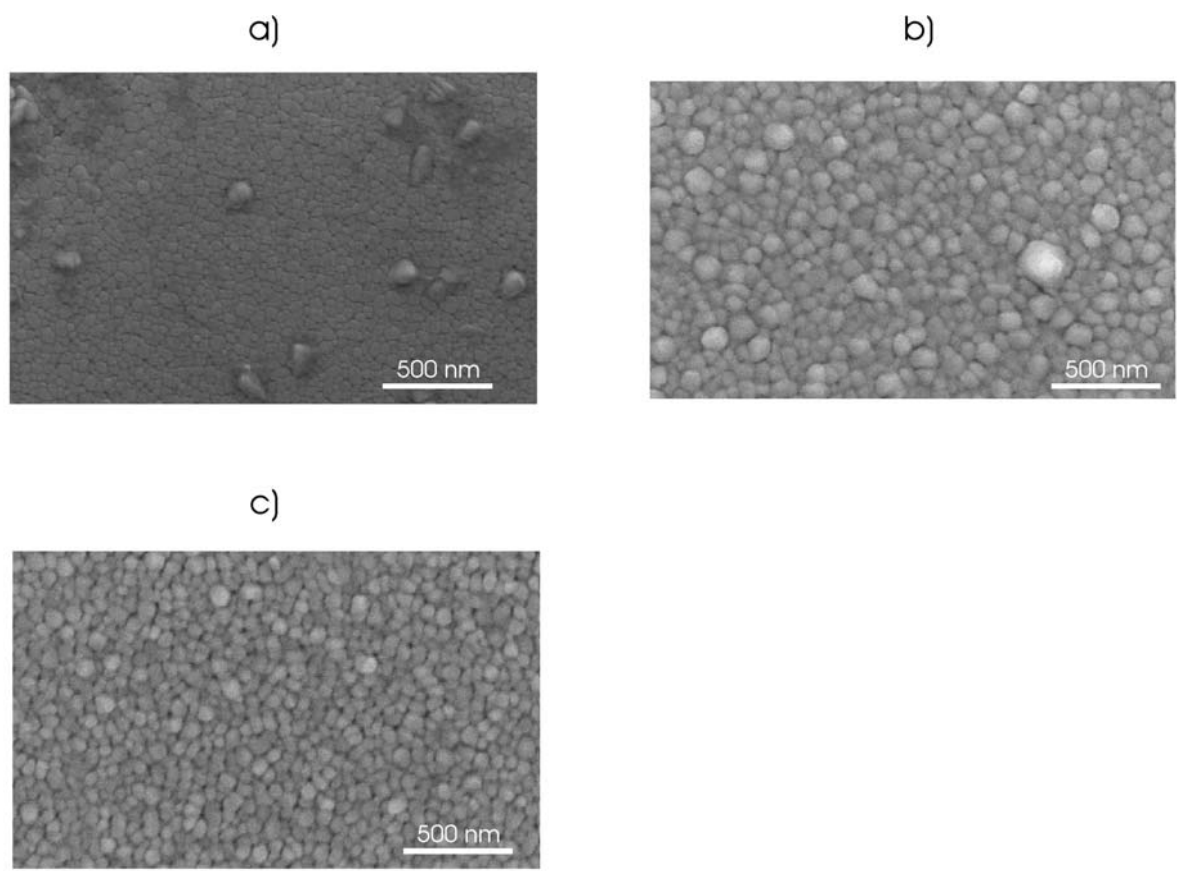

Fig. 2: SEM images of a LCMO film deposited on $SrTiO_3$ (a), $128°$cut $LiNbO_3$ (b) and YZ-cut $LiNbO_3$. The influence of the substrate on the film quality is clearly seen.

In Fig. 3 and Fig. 4, we depict the results of our X-ray investigations for the films of Fig. 2. A Theta/2Theta scan of LCMO deposited on $SrTiO_3$ (Fig. 3) indicated the expected epitaxial growth of LCMO. Besides the substrate peaks, only the (020) and (040) reflection of the LCMO was found.

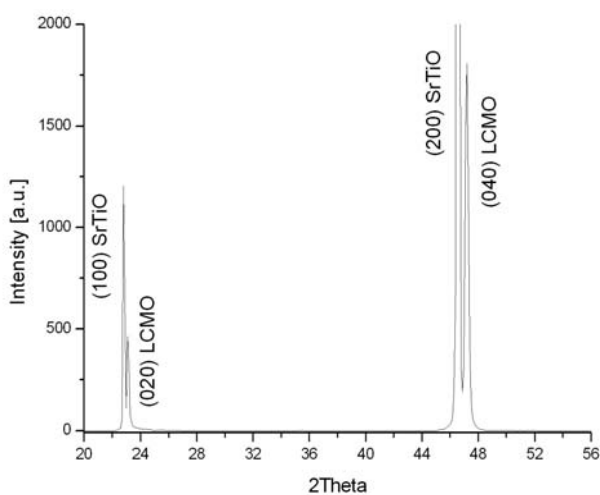

Fig.3: Theta/2Theta scan of LCMO deposited on $SrTiO_3$ indicating the epitaxial growth of LCMO on $SrTiO_3$

In Fig. 4, the corresponding pole figure for the two films of Fig. 2 (b) and 2(c) are shown. Whereas the pole figure of the LCMO film on YZ-cut $LiNbO_3$ (4b) exhibits no pregnant texture, the corresponding pole figure of the film deposited on 128°rot $LiNbO_3$ clearly shows reflections of different LCMO orientations. Thereby the pole figure was taken at a 2Theta value of 58.6° ((042) LCMO), in order to ensure that there is no recording of a substrate peak.

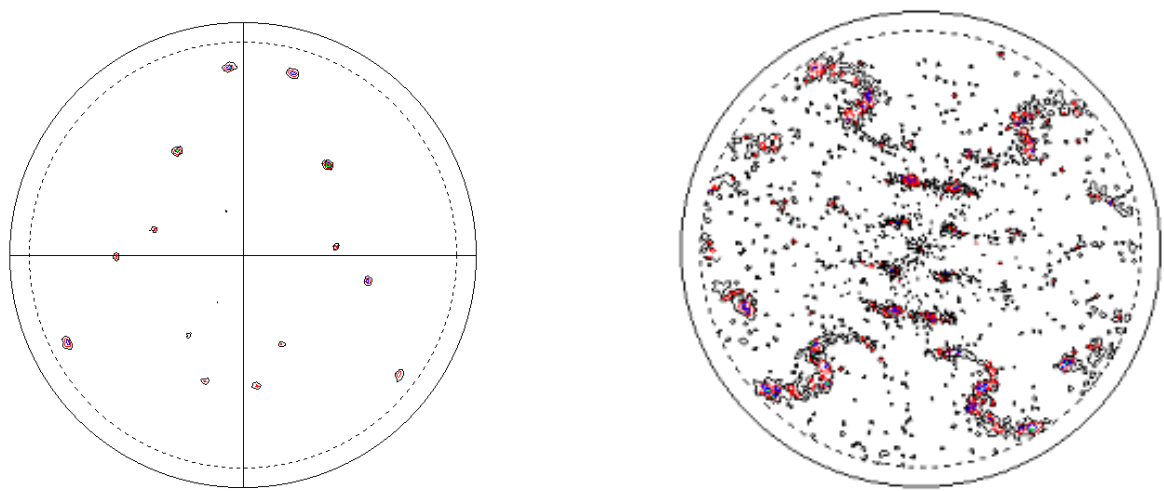

Fig. 4: Pole figure of LCMO deposited on 128°-rot $LiNbO_3$ (a) and YZ-cut $LiNbO_3$ (b). One clearly sees a higher degree of orientation in the case of 128°-rot $LiNbO_3$

Electrical characterization

For the first electrical characterization of the samples fabricated, we measured the four probe DC resistivity of the samples, using the Hall bar geometry as described above. The most striking feature in the transport behaviour of our Manganite thin films is the observation of a prominent temperature dependence of the resistivity as a function of temperature. This behaviour is generally interpreted in terms of a metal-insulator transition (MIT) [4]. On the SrTiO$_3$ substrate, this  metal insulator transition typically took place at around T$_{MIT}$=230K (FWHM 50K). On both types of the piezoelectric LiNbO$_3$ (128°-rot and YZ-cut) samples, we additionally measured the film resistance versus temperature curves at two different magnetic-fields (B=0 and B=9 T). As expected we found a strong, magnetic field induced decrease of the resistance and a pronounced shift to higher MIT-temperatures (as a measure of which we took the maximum in the ρ(T) data) with increasing magnetic field. The resistance on YZ-cut LNB at the MIT is about a  factor of 20 higher than on 128°-rot LNB, the MIT – temperature itself is shifted from about T$_{MIT}$=150 K on YZ-cut LNB to more then T$_{MIT}$=200 K on 128°-rot LNB. This shift has to be assigned to the influence of the different substrates. In both cases, the LNB substrates result in a MIT temperature being lower than on the SrTiO$_3$ substrate.

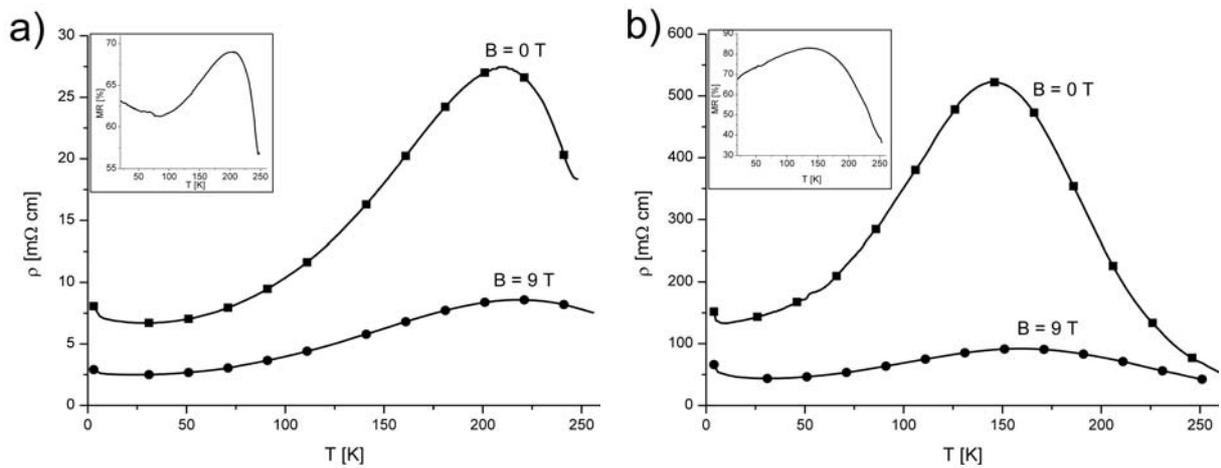

Fig. 5: Resistivity v. temperature at B=0 T (squares) and B=9 T (circles). In the inset the magneto resistance MR = (ρ[0T] – ρ[9T])/ρ[0T] vs. temperature is shown. a) on 128°-rot LNB; b) on YZ-cut LNB.

For comparison, and to gain insight into the integral resistivity distribution of the thin film, we simultaneously measured the temperature and magnetic field dependent SAW attenuation across the LCMO thin film.

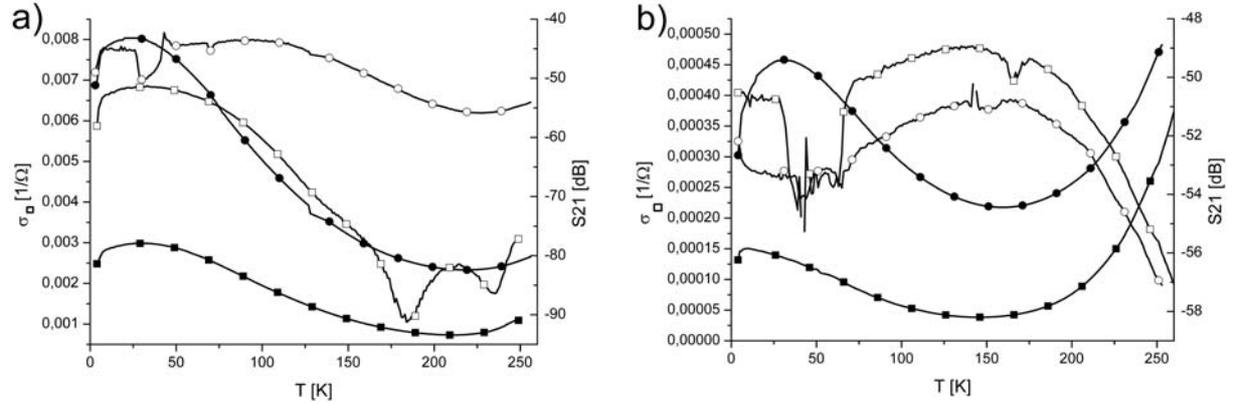

Fig. 6: Conductivity σ (0 T (closed circles) and 9 T(closed squares)) and SAW-attenuation Γ (0 T (open circles) and 9 T (open squares)) vs. temperature a) on 128°-rot LNB; b) on YZ-cut LNB.

Whereas the DC measurements provide resistivity data which might be influenced by micro cracks, film inhomogeneities, or alike, a SAW transmission measurement yields a contactless and more integral measurement of the sheet conductivity of the film. A SAW on a piezoelectric substrate acts on the film like a constant voltage source of short length ($\lambda_{SAW}$) and correspondingly large width ($W_{SAW}$). The piezoelectric interaction between the SAW and the sheet conductivity of a thin film is of relaxation type and turned out to be given by [2]:

$$\Gamma = \frac{K^2}{2} k \frac{\sigma/\sigma_m}{1+(\sigma/\sigma_m)^2} \qquad (1)$$

Here, σ is the sheet conductivity of the thin film, $K^2$ represents the electro mechanical coupling coefficient of the piezoelectric substrate, describing the coupling between mechanical and electrical part of the SAW. The wave vector $k=2\pi/\lambda_{SAW}$ accounts for the frequency dependent attenuation of the chip and $\sigma_m$ = 3e-7 $\Omega^{-1}_{sq}$ denotes the maximum of the conductivity dependent attenuation curve, according to eq. 1. If we plot the Γ(σ) curve as described by eq. 1, we end up with Fig. 7, where we show the dependence of the SAW attenuation as a function of the sheet conductivity of a thin conducting film on top of the substrate.

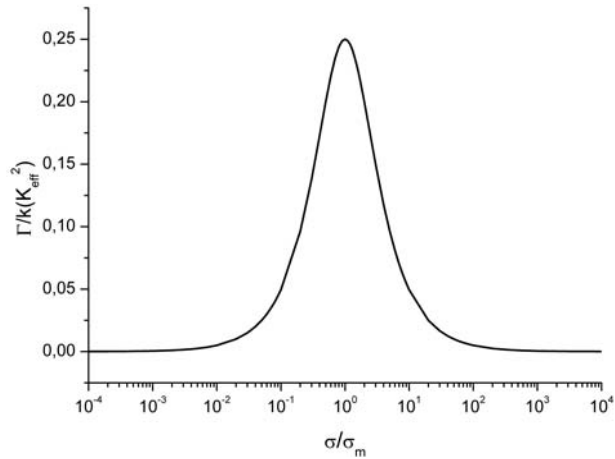

Fig. 7: Attenuation $\Gamma/k$ in units of $K_{eff}^2$ versus conductivity $\sigma$ in units of the characteristic conductivity $\sigma_m$.

Employing this well established relation between conductivity and SAW attenuation (eq. 1), we are able to calculate the expected SAW attenuation from the measured DC resitivity data. The result is shown by the solid lines in Fig. 8. Here, we plot the measured SAW attenuation $\Gamma(T, B)$ together with the calculated $\Gamma(T, B)$ curves as extracted from the DC measurements for both LNB samples. Clearly, an overall correspondence between the experiment and the calculated lines is observable, especially for the sample on 128°rot LNB (Fig. 8a). Here, the occurrence of the typical double peak structure in the zero field traces around T=200K proves the validity and applicability of our model. Such a double peak structure in the $\Gamma(\sigma)$ curve each time occurs, if, by undergoing a deep minimum, $\sigma(T, B)$ crosses $\sigma_m$ twice, once by falling below $\sigma_m$, and once by increasing again to $\sigma > \sigma_m$.

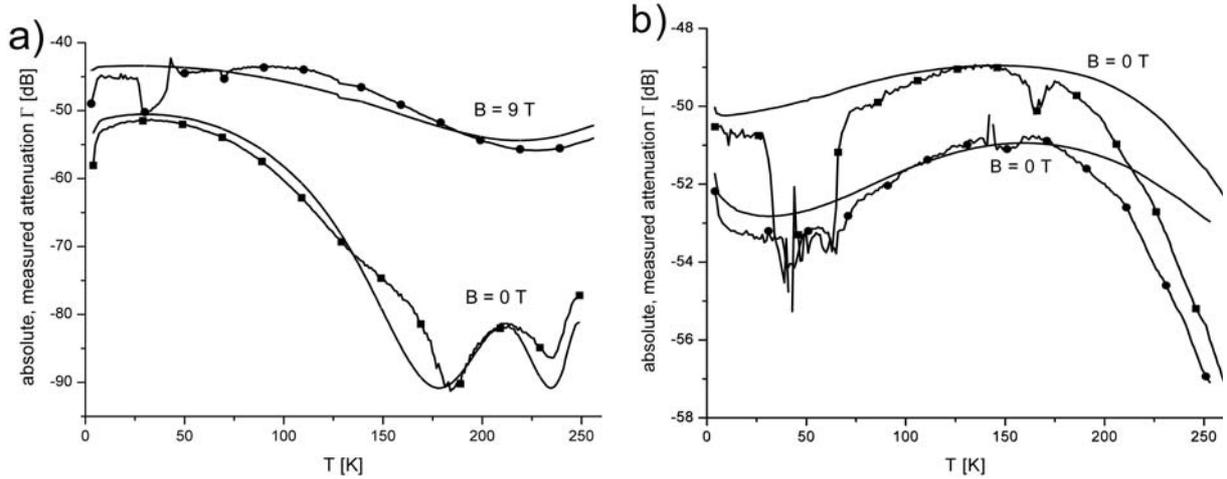

Fig. 8: Measured (closed symbols) and calculated (solid lines) attenuation Γ(T,B) for LCMO on LNB at two different magnetic fields. In (a), we show the results for 128°rot LNB substrate, in (b), the results for YZ cut LNB are shown. The overall agreement between measurement and calculation is clearly satisfactory.

Using eq. 1 we calculated the attenuation curves in Fig. 8 from the measured DC conductivity. To achieve best agreement, we fitted the traces by adjusting the free parameters $K_{eff}^2$ and $\sigma_m$, respectively. Here, the depth of the attenuation minimum is mainly determined by $K_{eff}^2$, whereas the spacing of the double peak minima (see Fig. 8a) is given by the critical conductivity $\sigma_m$ For the LCMO-film on 128°-rot LNB we obtain the fitting parameters $K^2 =$ 3.8 and $\sigma_m = 0.00025$ 1/Ω. As an additional correction term we had to introduce a small parasitic parallel conductivity $\sigma_{par} = 0.0006$ 1/Ω. For the LCMO-film on YZ-cut LNB the fitting parameters resulted in $K^2 = 0.4$ and $\sigma_m = 0.0009$ 1/Ω. Both values for $K^2$ are much higher as expected, the obtained values for $\sigma_m$ are in reasonable agreement with the expected values. The apparently very large $K_{eff}^2$ values are consistent with the measured extremely large absolute over all attenuation of the LCMO film on our samples. The shortness of the LCMO bridge in the direction of the SAW propagation would only result in a maximum attenuation of about $\Gamma_{max} = 0,2$ dB. However, we observe a total attenuation of up to 40 dB (cf. Fig. 8a).

Such a large overall attenuation can only be accounted for, if, for instance, an otherwise undetectable series conductivity is present on the sample. The average sheet conductivity of this parasitic series conductivity can add significantly to the measured attenuation, as the SAW interaction length in this case would dramatically increase as compared to the intentionally processed LCMO bar. In view of the fact, that in the first place the LCMO films have been deposited over the whole sample area between the IDTs, and then are removed by an etching process, we cannot exclude that a very thin, low conductivity film left remaining on the SAW delay line. Further studies will certainly clarify this point.

4. Conclusion

In summary, we have demonstrated the successful deposition of thin Manganite films on piezoelectric, lattice mismatched substrates employing a pulsed laser deposition technique. Temperature and magnetic field dependent conductivity measurements revealed a bulk like metal insulator transition of the thin film systems. The actual transition temperatures are dependent on the substrate chosen, being higher for substrates resulting in a better film morphology. Contact less surface acoustic wave transmission measurements on these films are used to yield information about the areal sheet conductivity of the films, and allow for a detailed investigation of the electrical properties of the films.

Acknowledgement

This work was supported by Deutsche Forschungsgemeinschaft DFG under the programme SFB 484.